# The processing goes far beyond "the app" – Privacy issues of decentralized Digital Contact Tracing using the example of the German Corona-Warn-App

Rehak, Rainer
*Weizenbaum-Institut for the Networked Society,
Berlin Social Science Center (WZB)
Berlin, Germany*
rainer.rehak@wzb.eu

Kühne, Christian R.
*Forum Computer Professionals
für Peace and Societal
Responsibility (FIfF)
Berlin, Germany*
christian.kuehne@fiff.de

**Since SARS-CoV-2 started spreading in Europe in early 2020, there has been a strong call for technical solutions to combat or contain the pandemic, with contact tracing apps at the heart of the debates. The EU's General Data Protection Regulation (GDPR) requires controllers to carry out a data protection impact assessment (DPIA) where their data processing is likely to result in a high risk to the rights and freedoms (Art. 35 GDPR). A DPIA is a structured risk analysis that identifies and evaluates possible consequences of data processing relevant to fundamental rights in advance and describes the measures envisaged to address these risks or expresses the inability to do so.**

**Based on the Standard Data Protection Model (SDM), we present the results of a scientific and methodologically clear DPIA. It shows that even a decentralized architecture involves numerous serious weaknesses and risks, including larger ones still left unaddressed in current implementations. It also found that none of the proposed designs operates on anonymous data or ensures proper anonymisation. It also showed that informed consent would not be a legitimate legal ground for the processing. For all points where data subjects' rights are still not sufficiently safeguarded, we briefly outline solutions.**

I. INTRODUCTION

*"It's not about privacy, it's about making a technology socially governable." Wilhelm Steinmüller (1934-2013)*

Especially in Europe, the discussion about the containment of the Corona pandemic has revolved around the use of technical tools, in particular the so-called corona tracing apps. These are supposed to automatically record epidemiologically relevant contact events of users and thus allow to warn and self-quarantine exposed contact persons to further prevent infections of individuals in a timely and retrospective manner.

Until the current Corona pandemic, contact tracing had been carried out manually in Europe by health authority employees, for example using the memory and calendars of infected persons and subsequently warning persons at risk by telephone. In some countries, such as China, many other sources of information and data categories are also used, such as credit card data or travel/movement information. This tedious work, so it was envisioned, could be assisted and therefore greatly accelerated through the use of digital contact tracing in the form of apps.

Even though the concrete suitability of such an app for this purpose is still controversial, both epidemiologically and technically [1][2], and there is a larger risk of a slow societal habituation to surveillance practices like permanent self-tracking and contact tracing, the focus in this paper is not the general question whether an app should be used at all, but how should such an app be designed in a data protection friendly manner. As a concrete example, the German Corona-Warn-App (CWA) [3] will be at the center of this article, because individual and societal consequences can only be meaningfully analyzed when looking at a concrete technical implementation. However the findings can be generalized for similar systems and could also be fed back into the concrete design and further development of the entire process in the form of legal, organisational or technical requirements.

*A. Data protection, IT security and privacy*

Data protection and its anchoring in legislation is a guarantor of fundamental rights and freedoms in the digital age, see Article 1 GDPR [4]. It relates not only to individual rights of data subjects, but also to collective rights such as the right to free assembly. More generally, data protection can also be said to maintain the functional differentiation of modern societies by problematizing structural asymmetries of information power [5] and thus safeguarding basic social functions and their inherent rationality. In summary, data protection has the goal to prevent unintended social and societal consequences of data processing [6] and therefore goes far beyond the individual [7][8]. Data protection concretely addresses this problem by protecting the data subjects and affected people from the processing organisation.

Data protection is therefore primarily concerned with preventing infringements of fundamental rights by the data processing itself and is only secondarily concerned with external attacks on systems or data and then only insofar as it is relevant to the data subjects. The focus of data protection is therefore not individual privacy, but the overall societal, structural and power effects of data processing [9]. Individual privacy might therefore be seen as (side) effect of good data protection.

In contrast, the goals of traditonal IT security are intended to protect the processing organization itself, its operations and its data from internal and external influence [10], even from influence by the data subjects themselves. The conflicts between data protection and IT security become apparent, although they are not the topic of this article.

Given the remarks above and on the contrary to IT security, a data protection analysis generally has to start with the processing organization as the primary source of risk [11] and only then it moves on to platforms, service providers, users and external third parties.

## II. Data protection impact assessment (DPIA)

In order to find, analyse and discuss the diverse consequences of this European project for large-scale contact tracing under government responsibility the European General Data Protection Regulation (GDPR) provides the instrument of a data protection impact assessment, which in specific cases is even mandatory. Concretely Art. 35 GDPR ("Data Protection Impact Assessment", DPIA) states:

*(1) Where a type of processing in particular using new technologies, and taking into account the nature, scope, context and purposes of the processing, is likely to result in a high risk to the rights and freedoms of natural persons, the controller shall, prior to the processing, carry out an assessment of the impact of the envisaged processing operations on the protection of personal data. A single assessment may address a set of similar processing operations that present similar high risks.*

*(7) The assessment shall contain at least*

*(a) a systematic description of the envisaged processing operations and the purposes of the processing, including, where applicable, the legitimate interest pursued by the controller;*

*(b) an assessment of the necessity and proportionality of the processing operations in relation to the purposes;*

*(c) an assessment of the risks to the rights and freedoms of data subjects referred to in paragraph 1; and*

*(d) the measures envisaged to address the risks, including safeguards, security measures and mechanisms to ensure the protection of personal data and to demonstrate compliance with this Regulation taking into account the rights and legitimate interests of data subjects and other persons concerned.*

It is essential for a DPIA according to the GDPR that the focus does not lie on the technology itself, in this case the Corona-Warn-App, but instead the DPIA should focus on the processing as a whole, which consists of several series of processing activities which can be, in part, supported by technology like an app. All considerations must therefore go beyond the use of "the app" and embrace the whole process including servers, network infrastructure or operating system frameworks and even the process parts without technology use. So to summarize it, the boundary of the app is not the boundary of the processing.

### A. Methodological approaches

There are different approaches for the methodological procedure of a DPIA. In Germany, the Conference of Federal and State Data Protection Commissioners generally recommends to use the "Standard Data Protection Model" (SDM [12]), which we will also utilise here, drawing from a full DPIA carried out by a group of researchers including ourselves [13]. This model first requires a threshold analysis to clarify the extent to which a DPIA for a given data processing system is not only societally desirable but also required by the law. Since contact tracing apps are a novel technology and process personal data on a large scale, and even medical data is processed in the case of infection, this is undoubtedly the case here.

The next step is to define the purpose of the entire data processing operation, in this case we exclusively look at the detection and interruption of infection chains. After that, it is important to work out the context of the processing. This includes not only the general social and political situation as well as technical circumstances, but also explicitly the different actors and their interests such as the German Ministry of Health ordering the CWA or its subordinated Robert Koch-Institut (RKI) operating it, but also the CWA users as well as people who explicitly do not want to use it or even employers thinking about using it as entrance requirement and even involved companies like Google and Apple. Only on this basis a well-founded analysis of risks and attack scenarios can be made later, including proportionality considerations.

Assumptions and use cases for the processing must then be developed and explicated in order to subsequently describe the processing activity en detail. It should be noted that the procedure must be broken down into sub-steps, not all of which have to be technology-supported as stated above. In the case here and as mentioned before, the procedure includes not only the app, but also the associated server systems, specialist applications and infrastructure components such as operating systems or technical communication relationships.

On this basis, the legal responsibility for the processing activity can then be assessed and legal requirements developed. Since the user can neither determine purpose nor means of the processing and cannot even configure anything within the app, the legally responsible party is solely the RKI, even if the CWA is called 'decentralized' [13].

Combining all this preliminary work, the vulnerabilities, hazards and risks of the processing can be mapped out. This means risks related to the fundamental rights of the data subjects and indeed all fundamental rights such as the right to free movement. Based on this risk analysis and the aforementioned legal requirements, concrete protective measures for the rights of the data subjects can then be determined and finally recommendations for the data controller must be listed. The recommendations should especially include the particularly problematic aspects, such as risks for which no protective measures exist.

## III. Architecture of the CWA

For reasons of minimizing the infringements of fundamental rights (and for simplifying the analysis), the referenced DPIA assumes a narrowly defined purpose for data processing: Simply warning individuals who have had contact with infected persons and are therefore at risk, which is implemented by the warning function of the CWA in Germany [3].

This warning functionality is put to practice by the smartphone sending regularly chaning strings (pseudonymous temporary identifiers, tempIDs) via Bluetooth at regular intervals using the "Bluetooth Low Energy Beacons" (BTLE) standard, and at the same time receiving the temporary identifiers (tempIDs) from other apps accordingly, when they are in close vicinity. Hence, each app keeps two buckets with the tempIDs of the last fourteen days, one bucket for the tempIDs sent and one for the tempIDs received. Actual location information, for example GPS data, is not processed or even collected by this system. The underlying Bluetooth functionality is managed by the Exposure Notification Framework (ENF or GAEN) provided by Apple and Google in their respective mobile operating systems [14].

In case of positive testing, only the temporary identifiers – the daily seeds, to be precise – sent out by the person during the past 14 days are uploaded to the CWA server. Those uploaded temporary identifiers indicate infectiousness. If any other app has received those tempIDs, it means that there was a possibly relevant contact event. Therefore the other apps regularly download the current data set of all infection-indicating tempIDs and check for matches, i.e. if they have seen any of those tempIDs. If yes, they calculate locally on the smartphone whether there is a risk of infection based on the duration and proximity of the contact, as well as the state of illness of the infected person at the time of contact. If there is a risk, the user is then warned accordingly. Since the server only knows the ever changing tempIDs of infected users, it can neither create a contact history nor calculate the social network of all users. The server does not even warn the users. Therefore, this decentralized variant is much more data protection-friendly, yet also more traffic-intensive than a centralised one, e.g. TousAntiCOVID in France or TraceTogether in Signapore.

## IV. KEY FINDINGS

In the following, we will present the key findings of the detailed DPIA [13] that are still relevant but also largely unsolved. The insights and learnings from this DPIA can also be generalised.

### A. Processing of personal health data and anonymising procedures

In the discussion about the features of a Corona app, there is often talk of an "anonymous system", and sometimes even the application of the GDPR itself is questioned because of this. However, as described in III. the procedure as a whole consists of processing contact data on smartphones, sending this data to a server after diagnosing an infection, and finally distributing it to all other smartphones to check for possible contacts with infected persons. All data on a smartphone is explicitly personal information, namely information related to the user of the device. This applies regardless of whether the device is well secured from the external parties or to what extent other apps can technically assign the received strings to a person. In principle it is always possible, because its a person's device [15].

Furthermore, because only those individuals who have been tested positively transmit data to the server, the transmitted data is even data concerning health, i.e. a 'special category' of personal data according to Art. 9 of the GDPR, which is particularly protected.

At this point, we see that the process as a whole is not anonymous and we should pay special attention to the required anonymization procedure when uploading the data to the server, so that at least on the server we don't have personal data, but only infection-indicating data. In an unprotected procedure, the operator of the server can directly establish a personal reference via the IP address of the uploading app, i.e. re-personalize the data and hence attribute the infection. However this is neither intended nor necessary for the given purpose and must therefore be prohibited to practise meaningful purpose limitation.

This personal reference should therefore be reliably separated from the tempIDs during the upload process so that the server can subsequently process and distribute them to other apps in a data protection friendly manner. It is precisely by this anonymising separation procedure, that the tempIDs become "infection-indicating data without personal reference" [13]. The identifiability relation is thus removed without changing the semantic content of the data itself or its usability for the intended purpose. This anonymization procedure can take many concrete forms and must be secured by a combination of legal, organizational and technical measures [16].

Legally, the operator must be independent and must not have any vested interests in the data. How much the research oriented RKI is independent and how much interests in the data it has must still be discussed. The RKI must also be protected from obligations to disclose data, including to government security agencies or other research organisations.

In organizational terms the responsible party, the German Ministry of Health, could strategically put in place two different operators: one operator, e.g. a public organisation, operates the input nodes to the network and hereby strips off the metadata, including the IP addresses, while the other one, e.g. the RKI, could operate the actual CWA server. Within the operators, attention must be paid to an appropriate departmental structure and separation of functions that enforce the separation of informational powers within the organization, i.e. to maintain functional differentiation [16].

Technically, the operator must implement the separation in such a way that uploads *cannot* be logged, neither on the server nor in their network, and, if necessary, make use of upstream anonymizing proxies (e.g. Tor). The connections must also be end-to-end encrypted.

All of the measures above must be made continuously verifiable and auditable through a data protection management system and must also be audited by the responsible local data protection officer and the data protection supervisory authorities.

### B. Voluntariness of use

The frequently asserted voluntariness of app use is quite a precondition-rich construct that may turn out to be an illusion in practice. For example, it is conceivable, that the use of the app could be considered a condition for individual relaxation of curfew restrictions. Presenting the app could serve as a condition to access public or private buildings, spaces, or events. Such a use would certainly not be covered by the purpose of the system, but could still be enforced by third-party actors (e.g., employers or private event organizers). This scenario would be an implicit coercion to use the app and would lead to a significant unequal treatment of non-users; the already existing "digital divide" between smartphone owners and non-owners would hereby expand to further areas of life. In addition, the purpose of the system could be undermined if users deliberately did not carry their smartphones with them on possibly dangerous trips (food shopping) fearing disadvantages or users could alternate between different devices. This risk can only be mitigated by accompanying legislation that effectively prevents those and other misappropriations.

### C. The problem of informed consent

The terms "voluntariness" and "consent" are often confused, especially concerning corona apps. The first term refers to whether people can decide to use the app themselves or whether the use of the app is mandatory, i.e. it must be used by everyone. The second term, consent, is specifically concerned with the legal basis under data protection law on which data processing is to take place (Art. 6 GDPR). In the case of consent (Art. 6 (1) lit. a GDPR), the user is presented with the information what data processing the app and the system behind it will perform and then confirmation is obtained for it. Insofar as this decision is made in an informed, specific, active and uninfluenced manner pursuant to Art. 4 (11) GDPR, it is legally considered consent (Art. 6 (1) lit. a, Art. 7 GDPR). However, there are also other legal bases for data processing, such as the necessity of data processing for the fulfillment of a contract (Art. 6 (1) (b) GDPR) or on the basis of a law (Art. 6 (1) (e) GDPR).

The voluntary nature of the app use (see IV. B.) must be seen independently of the legal basis for the procedure as a whole under data protection law. While there can be no compulsory app with consent, a voluntary app with a legal basis is conceivable and even desirable. It is desirable because in a consent solution, the party responsible for the procedure and therefore the operator of the app alone decide what exactly the users consent to.

Particularly in the present constellation, in which the app on the smartphone operates exactly as determined by the responsible party and exclusively under the control of this party, comparable to a DRM system [17], weighing the fundamental rights risks against achieving the purpose is externalized to the data subjects [11]. Hence, the informational power asymmetry between the responsible party and the data subject caused by the processing is thus not balanced by the consent mechanism, but reinforced [7][13]. Furthermore, with the RKI as controller, a German higher federal authority, it is anyhow highly questionable how freely individuals can consent to such a citizen-state relationship.

On the other hand, in the case of a voluntary app based on a law, the directly democratically legitimized legislature would explicitly determine the processing and also define its limits in a law based on democratic negotiation. For this purpose, an extension of the German Infection Protection Act (IfSG), for example, could be considered. In addition, it should be noted that the GDPR allows extensive use of existing data, e.g. CWA server data, for research purposes according to Art. 89 GDPR. If there is a legal basis for data processing, the exact purposes can be made explicit for research as well, or such further uses can be explicitly prohibited after a negative proportionality test.

Against the background of these considerations, a voluntary app based on a legal foundation is both desirable from a data protection perspective and also legally required. In addition, a further legal provision would have to prohibit the misuse of the app for access management for buildings or workplaces, as stated in III. B. On the one hand, this is necessary to make the use of the app not only de lege but also de facto voluntary, and on the other hand to support the actual achievement of the purpose. Because if the purpose can not be properly achieved by the app system, it would no longer be suitable and therefore no longer permissible.

However, up to today, the German CWA still uses the consent mechanism to suffice the needs of the GDPR [3].

### D. Ensuring the ability of data subjects to intervene

Without the ability to intervene in the data processing and a tight purpose limitation, the protection of fundamental rights is at risk: There is a high risk of falsely registered exposure events (false positives through walls, situations with masks and ventilation or laboratory errors), which would result in unjustly imposed self-quarantine. Countering this requires legal and factual means of effective influence, such as recalling false infection reports, deleting falsely registered contact events, or challenging possible other consequences. No known app including the CWA technically allows this so far.

### E. The role of platforms and software infrastructure providers

The role of the mobile operating systems manufacturers and app platform providers Apple (iOS) and Google (Android) must be critically discussed and accompanied throughout the entire processing procedure. A Bluetooth-based corona tracing app relies on the cooperation of the platform providers for technical reasons, as this kind of access to the Bluetooth module of the devices must be enabled at the operating system level.

Last year, the platform providers have used their position of technological and infrastructural power to enforce a decentralized and thus more data protection-friendly architecture like DP3T [18] against the intention of numerous governments who wanted to deploy centralized solutions. From a data protection point of view this is desirable in terms of the outcome, but highly problematic in terms of the political process. In addition, this move has largely lost sight of the data protection risk posed by the platform operators themselves in the public discussion.

As an operating system manufacturer, it is possible in principle (and also realistic, as the DPIA shows) for Google and Apple to obtain the contact information and derive information about infection cases and exposure risks from it. Moreover, the source code of the ENF/GAEN is still kept secret, even to the data protection supervisory authorities. Continuous critical monitoring of the role of Apple and Google therefore requires comprehensive awareness of this problem and the not only legal but factual obligation of the companies to behave in a data protection-compliant manner.

### V. CONCLUSION

With the data protection impact assessment of the CWA summarized here, a methodological sound form for analyzing, determining, and, if necessary, mitigating the data protection risks of a tracing apps has been presented. A generalization of the approach used here can surely also be utilised for analysing other data processing procedures vulgo IT systems affecting data subjects and their fundamental rights and freedoms.

In particular this assessment shows two things, that are still very relevant for the academic and the public discourse regarding digital contact tracing: First, data protection and IT/data security are very different fields, although there are points of contact. This difference expresses itself in terms of goals and also in terms of measures. This difference becomes especially relevant in data protection impact assesments and the origin of risks.

Secondly, the open-source development of servers and apps along with all their components – for example, as free software – is an essential prerequisite for ensuring the necessary transparency regarding the implementation of data protection principles not only for data protection supervisory authorities, but also for those affected and the public at large. However, this paper and the referenced data protection impact assessment also shows that a technical focus on the openness and transparency of the software source code alone can obscure the larger social and societal implications of the entire process. In addition, the focus on open-sourced server and app development obscured the fact, that much of the infrastructure used by the app, namely the CWA relevant parts in the mobile operating system software of Google and Apple have still not been checked by any authority.

Only high-quality data protection impact assessments can reveal such implications and can lay a foundation for wider societal debates and informed political negotiations about the use and role of information technology. Proper DPIAs can help to enable and improve such discourses in a pluralistic and democratic society. Therefore DPIAs should generally be published so that they can be discussed not only by data protection regulators, but also by researchers, journalists, civil society and the general public [19].

ACKNOWLEDGMENT

We deeply thank Kirsten Bock, Rainer Mühlhoff, Měto R. Ost and Jörg Pohle for valuable input for this paper and fruitful discussions about the nature of data protection.